\documentclass[11pt,a4paper]{article}

\usepackage[margin=1in]{geometry}
\usepackage{ifthen}
\usepackage{xcolor}
\usepackage{graphicx}
\usepackage{algorithm}
\usepackage{algpseudocode}
\usepackage{booktabs}
\usepackage{tikz}
\usepackage{amsmath,amssymb}
\usepackage{microtype}
\usepackage[colorlinks=true,linkcolor=blue,citecolor=blue,urlcolor=blue]{hyperref}

\graphicspath{{./images/}{./}}
\newboolean{gdanonymous}\setboolean{gdanonymous}{false}
\newcommand{\libname}{MSAGLJS}
\newcommand{\msagljs}{\textsf{\libname}}

\newcommand{\anonsource}{https://github.com/microsoft/msagljs}
\newcommand{\canonsource}{https://github.com/microsoft/msagljs}
\newcommand{\anondemowebgl}{}
\newcommand{\anonurl}[1]{\url{#1}}
\newcommand{\anondemourl}[2]{\url{#2}}

\providecommand{\supplement}[1]{}

\newcommand{\comm}[1]{}

\newcommand{\cdt}{$\mathcal{T}$}

\hypersetup{
  pdftitle={Browsing Large Graphs with Tile Pyramids and Sleeve Routing in the Browser},
  pdfauthor={Lev Nachmanson, Xiaoji Chen}
}

\title{Browsing Large Graphs with Tile Pyramids and\\
       Sleeve Routing in the Browser}
\author{%
  Lev Nachmanson\thanks{Microsoft Research, Redmond, US.\
    \texttt{levnach@hotmail.com}}%
  \and
  Xiaoji Chen\thanks{OpenAI.\ \texttt{cxiaoji@gmail.com}}%
}
\date{\today}

\begin{document}

\maketitle

\begin{abstract}
We present a new way to visualize a large graph in the style of
online geographic maps. The method builds a \emph{tile pyramid} for
\emph{semantic zoom}: at every zoom level the labels of the
highest-ranked nodes remain readable, just as the names of major
geographical features stay readable on those maps.

The edges are routed by a method we call \emph{sleeve routing}, which
searches the dual graph of a Constrained Delaunay Triangulation to
select a sequence of triangles through the free space, then applies the
funnel algorithm to compute a shortest path inside the selected
sleeve. We apply several heuristics to speed up the routing.

We implemented our approach in the WebGL renderer of \msagljs{}\ifthenelse{\boolean{gdanonymous}}{\footnote{GJS is a placeholder name used during double-blind review; the actual library name will be disclosed in the camera-ready version.}}{}, an
open-source TypeScript library for graph visualization in web
browsers, with the entire pipeline running client-side, without a
dedicated server. Our benchmark suite contains nine graphs with up to
$32{,}768$ nodes and $236{,}978$ edges, and measures browser-side
parsing, layout, routing, and tile-pyramid construction. The renderer's
demo can be seen at
\anondemourl{\anondemowebgl}{https://microsoft.github.io/msagljs/renderer-webgl-sleeve/index.html}.

\msagljs{} is available
\ifthenelse{\boolean{gdanonymous}}{as open source\footnote{Anonymized source code mirror for review: \url{\anonsource}}}{on GitHub\footnote{\url{\canonsource}}} and as
NPM packages\footnote{\ifthenelse{\boolean{gdanonymous}}{Package names omitted for double-blind review.}{\texttt{@msagl/core}, \texttt{@msagl/drawing},
\texttt{@msagl/parser}, \texttt{@msagl/renderer-svg},
\texttt{@msagl/renderer-webgl} --- all on
\url{https://www.npmjs.com/}.}}.
\end{abstract}

\section{Introduction}
\label{sec:intro}

Visualizing graphs in web browsers without a dedicated server presents
unique challenges: the runtime is single-threaded, the per-tab
JavaScript heap is capped at roughly 4\,GB,\footnote{In current
Chrome, V8 with pointer compression caps the per-tab JavaScript heap
at approximately $4\,\text{GB}$ ($2^{32}$ bytes).} and users expect
the smooth pan-and-zoom experience familiar from online
maps. \msagljs{}
is an open-source TypeScript library that addresses these challenges
by combining efficient layout algorithms, a novel edge routing method,
and a tiling scheme with semantic zoom.

\msagljs{} is consumed as a set of NPM packages and renders graphs
using WebGL through the deck.gl framework~\cite{deckgl}. It supports
the layered Sugiyama scheme~\cite{sugiyama1981}, Pivot
MDS~\cite{brandes2007eigensolver}, and
IPSep-CoLa~\cite{dwyer2006ipsepcola}, with edges routed around node
obstacles. Throughout this paper our typical example is a social
network laid out by IPSep-CoLa; the methods we describe, however, are
agnostic to the layout algorithm and apply to any node placement in
which the nodes do not overlap. The graph loading pipeline builds a
hierarchy of tiles---analogous to map tiles---so that at any zoom
level only a bounded number of entities are drawn. Tiles are delivered
to the browser through the standard \texttt{TileLayer} of the
\texttt{@deck.gl/geo-layers} package, giving smooth pan and zoom. A live
demo of the WebGL renderer is available
online\footnote{\anondemourl{\anondemowebgl}{https://microsoft.github.io/msagljs/renderer-webgl-sleeve/index.html}}.

This paper makes two main contributions:
\begin{enumerate}
\item A \emph{tiling scheme} for large graph visualization that builds a hierarchical tile pyramid, limits visible entities per tile using PageRank-based filtering, simplifies edge routes at coarser levels, and integrates with the WebGL renderer for real-time pan and zoom.
\item A \emph{sleeve routing} algorithm that routes edges on the dual graph of a Constrained Delaunay Triangulation, using per-source batched Dijkstra trees followed by the classical funnel algorithm to produce shortest-path geodesics in homotopy classes.
\end{enumerate}

\section{Related Work}
\label{sec:related}

\paragraph{Graph visualization tools.}
Graphviz~\cite{gansner2000graphviz,graphviz} is a long-standing graph
drawing tool supporting multiple layout algorithms, including the
Sugiyama method and Scalable Force-Directed Placement~\cite{sfdp}; it
does not support tiling or WebGL rendering. \ifthenelse{\boolean{gdanonymous}}{MSAGL~\cite{nachmanson2004glee} is a
desktop layout engine that did not have the features described here.}{\msagljs{} builds on the
earlier MSAGL desktop layout engine~\cite{nachmanson2004glee} that did
not have the features described here.} On the web,
Sigma.js~\cite{sigmajs} and Cytoscape.js~\cite{franz2016cytoscape}
provide interactive graph rendering, but rely on straight-line or
generic spline edges without obstacle-avoiding routing and without a
pyramid of precomputed tiles. ReGraph~\cite{regraph} uses WebGL to
render large graphs with straight-line edges and supports interactive
cluster expansion but not tiling. Cosmograph~\cite{cosmograph} uses
GPU-accelerated force-directed layout for million-node graphs with
straight-line edges, without
tiling. GraphMaps~\cite{nachmanson2015graphmaps} introduced the
tile-pyramid and node-ranking scheme\ifthenelse{\boolean{gdanonymous}}{}{ on which we build}, but it runs
only on Windows, ships raster tiles, and reuses one global edge
routing that is merely clipped per tile; \msagljs{} brings the idea to
the browser with vector tiles and per-level sleeve routing. Cornac~\cite{perrot2018cornac} handles huge
graphs with tiles and edge
bundling~\cite{hurter2012graph,holten2009fdeb} but requires a
multi-server backend. Earlier navigation schemes for large graphs
include ASK-GraphView~\cite{abello2006askgraphview}, which organizes
nodes into a hierarchical cluster tree, and topological fisheye
views~\cite{gansner2005topological}, which distort a multilevel
layout~\cite{harel2002multiscale,hu2005efficient} under the focus;
neither targets browser-side tile pyramids or per-level edge
routing. Hierarchical edge bundling~\cite{holten2006hierarchical}
and force-directed bundling~\cite{holten2009fdeb} are complementary
clutter-reduction techniques applied \emph{after} edges have been
routed.

\paragraph{Shortest paths and edge routing.}
Our sleeve algorithm walks the dual of a Constrained Delaunay
Triangulation~\cite{shewchuk2002delaunay} and relies on the classical
funnel algorithm for extracting a geodesic from a triangle
strip~\cite{lee1984euclidean,chazelle1982theorem,guibas1987linear,pathOpt}.
The optimal algorithm for Euclidean shortest paths among polygonal
obstacles~\cite{hershberger1999optimal} and the recent
$O(m\log^{2/3} n)$ single-source shortest-path algorithm that breaks
the sorting barrier~\cite{duan2025sorting} are remarkable theoretical
results, but they are not practical: both build elaborate global data
structures that we avoid.
Graphviz builds the full visibility graph for edge routing in force-directed layouts, which has $O(n^2)$ edges. For Sugiyama layouts, it uses the funnel algorithm~\cite{dobkin1997implementing} but requires a simple containing polygon. yWorks~\cite{yworks} offers ``Organic edge routing'' based on force-directed simulation. The approach of Dwyer and Nachmanson~\cite{dwyer2010fast} routes on a Yao-graph spanner with local shortcutting. For orthogonal layouts, the libavoid framework of Wybrow, Marriott, and Stuckey performs obstacle-avoiding connector routing with incremental updates~\cite{wybrow2010orthogonal}.

\paragraph{CDT-based pathfinding.}
The closest precedent to our routing pipeline is the CDT-and-funnel
pathfinder of Demyen and Buro~\cite{demyen2006efficient}, originally
designed for game-AI units of nonzero radius. They run admissible A*
search on the CDT dual (with a reduced-graph variant that collapses
corridor chains into single abstract edges) to recover provably
shortest paths. Their search is, however, per-query: the bounds depend
on the start and goal, so it cannot share work across the edges with
the same source as our per-source Dijkstra does. For the graphs we
target this would be roughly an order of magnitude slower; we
therefore trade their proven optimality for speed.

\section{Tiling}
\label{sec:tiling}
The input to tiling is a laid-out graph together with routed edges: we
prefer to route the edges with sleeve routing,
Section~\ref{sec:sleeve-routing}, to achieve visual stability across
level changes. The output is a sequence of $Z{+}1$ levels: level~$Z$
is the finest, and level~$0$ is the coarsest. Each level is a
collection of tiles indexed by integer $(x,y)$ grid coordinates.

A tile is a rectangular viewport together with its \emph{tile data}:
nodes, edge labels, edge arrowheads, and \emph{edge clips}. An edge
clip is a curve confined to the tile rectangle, paired with the
non-empty list of graph edges whose routes enter and exit the tile at
very close endpoints. Bundles usually capture the stretches on which
several routes travel close together inside a common sleeve; we
represent each such stretch as a single clip with an attached edge
list, rather than as one clip per edge.

Pseudocode~\ref{alg:build-pyramid} describes the algorithm building
the tile pyramid.
\begin{algorithm}[t]
\caption{\textsc{BuildTilePyramid}$(G, \mathit{TileCapacity}, \mathit{minTileSize}, \mathcal{M}_{\max})$}
\label{alg:build-pyramid}
\begin{algorithmic}[1]
\State $T_0 \gets$ single tile whose rectangle is the smallest axis-aligned square of side $2^{\lceil \log_2 \max(w,h)\rceil}$ centered on $G$'s bounding box of width $w$ and height $h$, holding all of $G$
\State $\mathit{levels} \gets [\,T_0\,]$;\quad $z \gets 0$;\quad $\mathit{used} \gets |T_0|$ \Comment{$|T|$ = element count of tile $T$}
\While{some tile in $\mathit{levels}[z]$ exceeds $\mathit{TileCapacity}$ elements}
  \State $\mathit{next} \gets$ empty $2^{z+1}{\times}2^{z+1}$ tile grid
  \ForAll{tiles $T \in \mathit{levels}[z]$} \Comment{per-tile inner loop}
    \State split $T$ into its four sub-tiles of $\mathit{next}$ \label{algln:split}\Comment{Section~\ref{sec:tile-build}}
    \State $\mathit{used} \mathrel{+}=$ number of new elements added to $\mathit{next}$
    \If{$200 \cdot \mathit{used} > \mathcal{M}_{\max}$} \Comment{memory budget; checked mid-split}
      \State \textbf{discard} $\mathit{next}$ and \textbf{break} pyramid growth
    \EndIf
  \EndFor
  \If{$\mathit{next}$'s tile width or height is below $\mathit{minTileSize}$}
    \State \textbf{break} \Comment{tile size below threshold}
  \EndIf
  \State $\mathit{levels}[z{+}1] \gets \mathit{next}$;\quad $z \gets z + 1$
\EndWhile
\State $Z \gets z$;\quad $V_Z \gets V(G)$;\quad $s_v^{(Z)} \gets 1$ for all $v\in V_Z$
\State $V_{\mathrm{PR}} \gets$ nodes of $G$ sorted by descending PageRank
\For{$z = Z-1$ \textbf{down to} $0$}
  \State $(V_z, \{s_v^{(z)}\}) \gets$ \textsc{SelectTopKWithAdaptiveScale}$(V_{\mathrm{PR}},\, \lceil |V|/2^{Z-z}\rceil)$ \Comment{Section~\ref{sec:filter}}
  \State build a CDT from the inflated polylines of $V_z$ at scales $\{s_v^{(z)}\}$
  \State route every edge $e\in E_z$ on this CDT
  \State rebuild the tiles of $\mathit{levels}[z]$ \label{algln:rebuild}\Comment{Section~\ref{sec:tile-build}}
\EndFor
\State \Return $\mathit{levels}$
\end{algorithmic}
\end{algorithm}

\begin{figure}[!tb]
  \centering
  \resizebox{0.7\linewidth}{!}{%
  \begin{tikzpicture}[
      x={(1.0cm,0cm)},
      y={(0.40cm,0.60cm)},
      z={(0cm,1.0cm)},
      line cap=round,
    ]
    \def\W{2.0}      
    \def\dh{1.4}     
    \def\cx{-0.25} \def\cy{0}
    \pgfmathsetmacro{\iwfull}{2*\W}
    \foreach \zlev/\h in {%
        4/0,%
        3/1,%
        2/2,%
        1/3,%
        0/4%
      } {
      \pgfmathtruncatemacro{\nn}{int(2^\zlev)}
      \pgfmathsetmacro{\step}{2*\W/\nn}
      \pgfmathsetmacro{\twov}{pow(2,\zlev-1)}
      \pgfmathsetmacro{\hx}{1226/(512*\twov)}
      \pgfmathsetmacro{\hy}{950/(512*\twov)}
      \filldraw[fill=blue!4, draw=blue!55, line width=0.5pt]
        (-\W,-\W,\h*\dh) -- (\W,-\W,\h*\dh) -- (\W,\W,\h*\dh) -- (-\W,\W,\h*\dh) -- cycle;
      \pgfmathsetmacro{\iw}{2*\hx}
      \pgfmathsetmacro{\ih}{2*\hy}
      \pgfmathsetmacro{\vxa}{\cx-\hx}
      \pgfmathsetmacro{\vya}{\cy-\hy}
      \begin{scope}
        \clip (-\W,-\W,\h*\dh) -- (\W,-\W,\h*\dh) -- (\W,\W,\h*\dh) -- (-\W,\W,\h*\dh) -- cycle;
        \begin{scope}[shift={(\vxa,\vya,\h*\dh)}]
          \pgflowlevelsynccm
          \pgftransformcm{1}{0}{0.40}{0.60}{\pgfpointorigin}
          \pgftext[base,left]{%
            \ifcase\zlev
              \includegraphics[width=\iw cm, height=\ih cm]{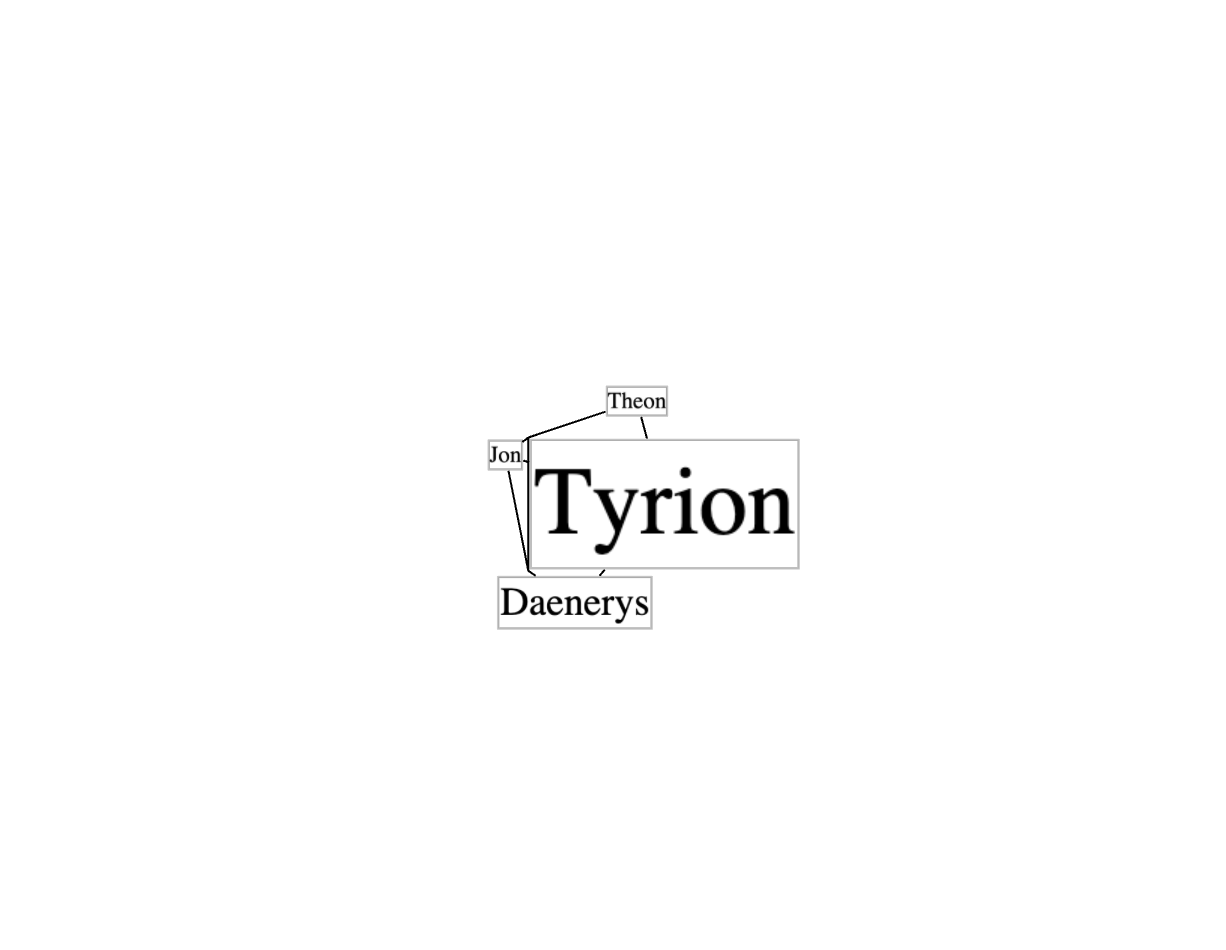}\or
              \includegraphics[width=\iw cm, height=\ih cm]{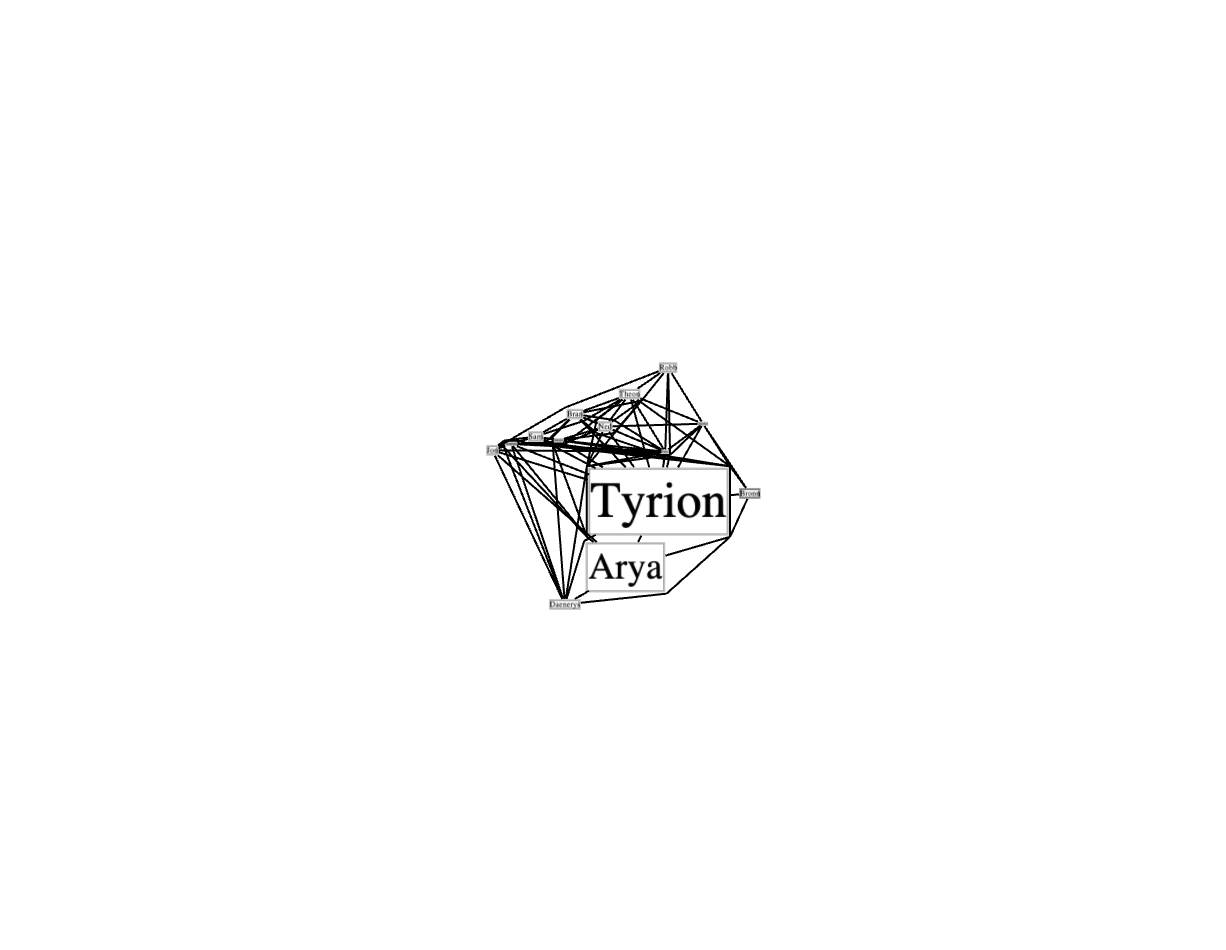}\or
              \includegraphics[width=\iw cm, height=\ih cm]{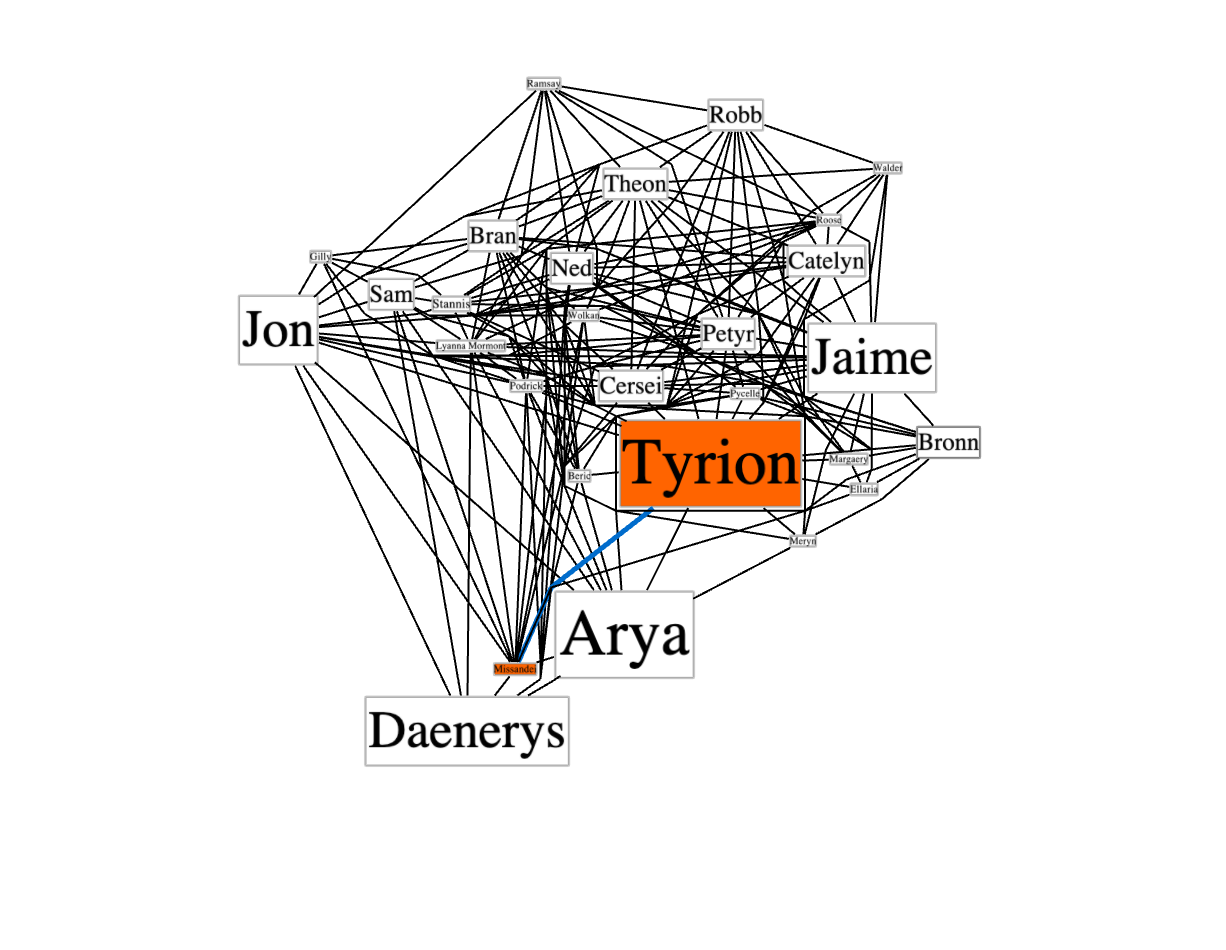}\or
              \includegraphics[width=\iw cm, height=\ih cm]{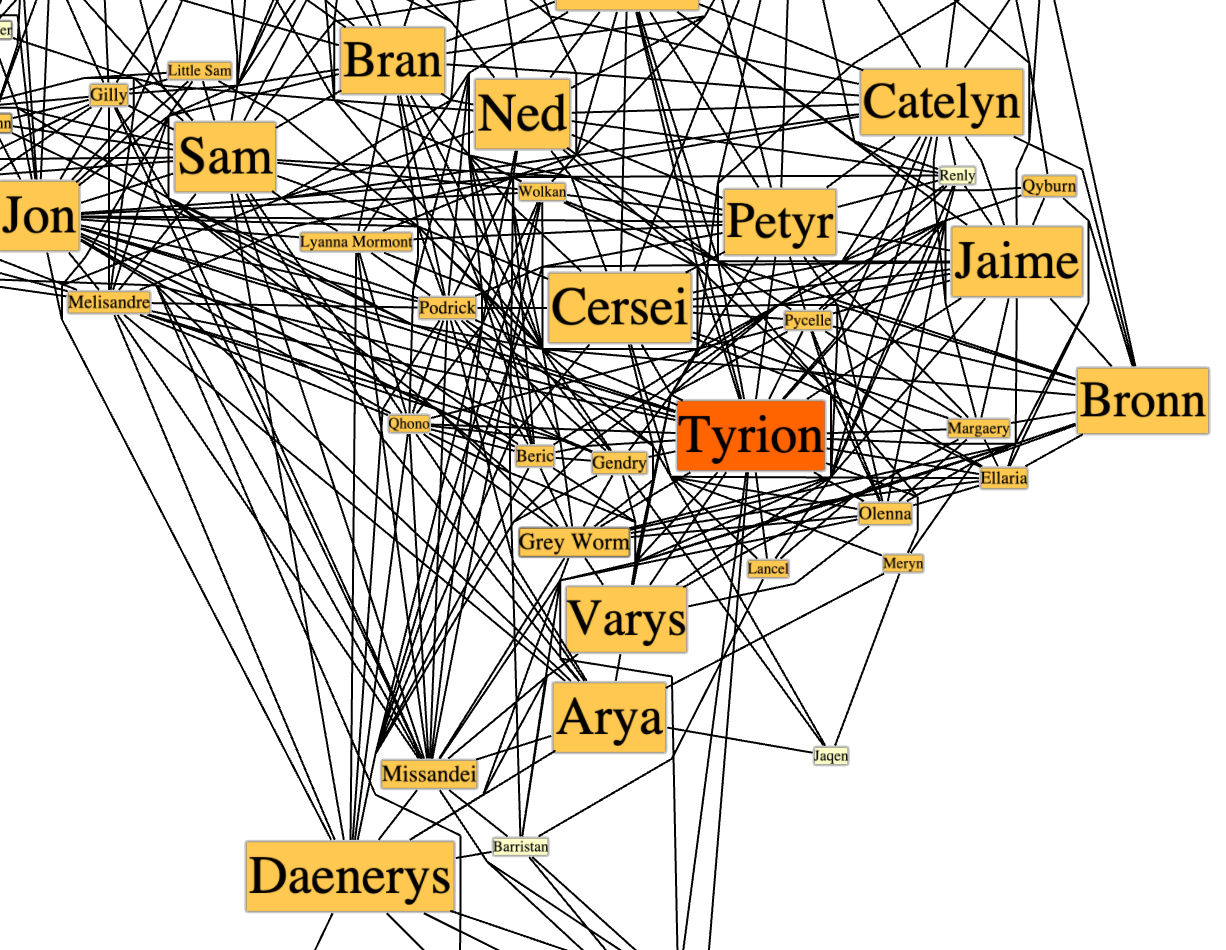}\or
              \includegraphics[width=\iw cm, height=\ih cm]{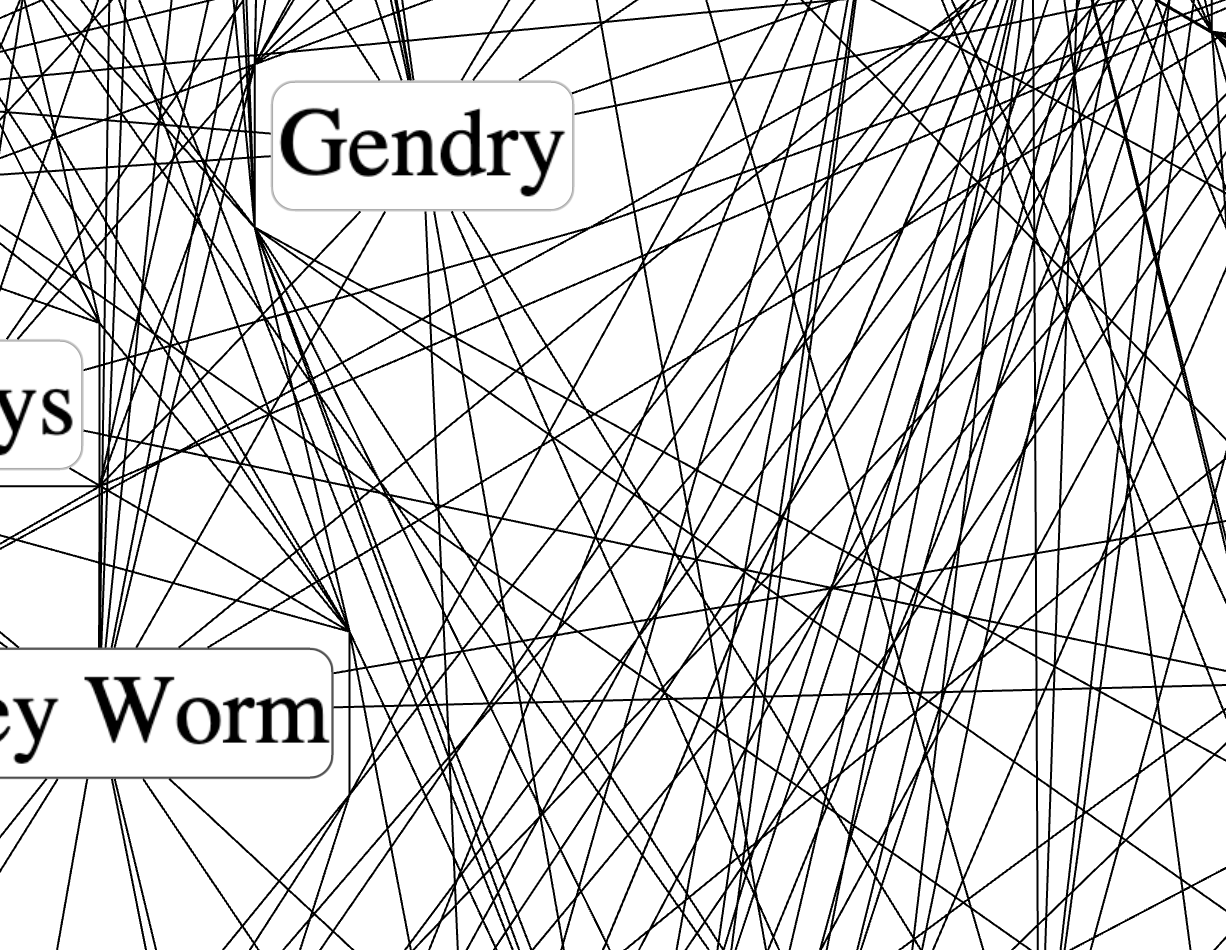}%
            \fi}
        \end{scope}
      \end{scope}
      \ifnum\nn>1
        \foreach \i in {1,...,\numexpr\nn-1\relax} {
          \pgfmathsetmacro{\u}{-\W + \i*\step}
          \draw[blue!50, line width=0.25pt] (\u,-\W,\h*\dh) -- (\u,\W,\h*\dh);
          \draw[blue!50, line width=0.25pt] (-\W,\u,\h*\dh) -- (\W,\u,\h*\dh);
        }
      \fi
      \ifnum\zlev=3
        \pgfmathsetmacro{\vxa}{\cx-\hx} \pgfmathsetmacro{\vxb}{\cx+\hx}
        \pgfmathsetmacro{\vya}{\cy-\hy} \pgfmathsetmacro{\vyb}{\cy+\hy}
        \pgfmathtruncatemacro{\ia}{floor((\vxa+\W)/\step)}
        \pgfmathtruncatemacro{\ib}{floor((\vxb+\W)/\step)}
        \pgfmathtruncatemacro{\ja}{floor((\vya+\W)/\step)}
        \pgfmathtruncatemacro{\jb}{floor((\vyb+\W)/\step)}
        \foreach \i in {\ia,...,\ib} {
          \foreach \j in {\ja,...,\jb} {
            \pgfmathsetmacro{\xa}{-\W + \i*\step}
            \pgfmathsetmacro{\xb}{-\W + (\i+1)*\step}
            \pgfmathsetmacro{\ya}{-\W + \j*\step}
            \pgfmathsetmacro{\yb}{-\W + (\j+1)*\step}
            \draw[orange!75!black, line width=0.7pt]
              (\xa,\ya,\h*\dh) -- (\xb,\ya,\h*\dh) -- (\xb,\yb,\h*\dh) -- (\xa,\yb,\h*\dh) -- cycle;
          }
        }
      \fi
      \node[anchor=east, font=\scriptsize] at (-\W-0.05, -\W, \h*\dh)
        {$z{=}\zlev$\,(\nn${\times}$\nn)};
    }
  \end{tikzpicture}}
  \caption{The entities rendered on each level are shown without
    scaling as the user zooms in.}
  \label{fig:tile-pyramid-3d}
\end{figure}
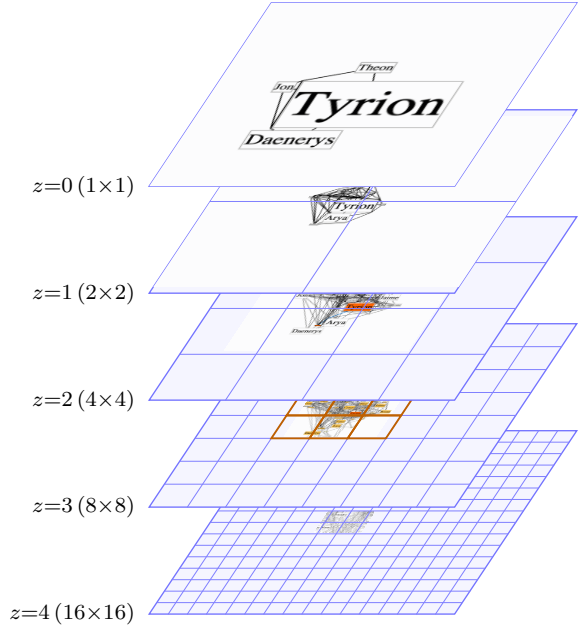

\paragraph{How the index of the finest layer, $Z$, is chosen.}
We treat the tiles differently when calculating $Z$ and when preparing
for rendering. There is no element filtering in the former case: all
graph entities are represented in each level. We continue increasing
$Z$, generating new levels, and splitting tiles, until a stopping
condition is met. Then, while keeping the finest level intact, we
rebuild the tiles of levels~$0, \ldots, Z-1$ for rendering.

\begin{figure}[t]
  \centering
  \includegraphics[width=0.48\textwidth]{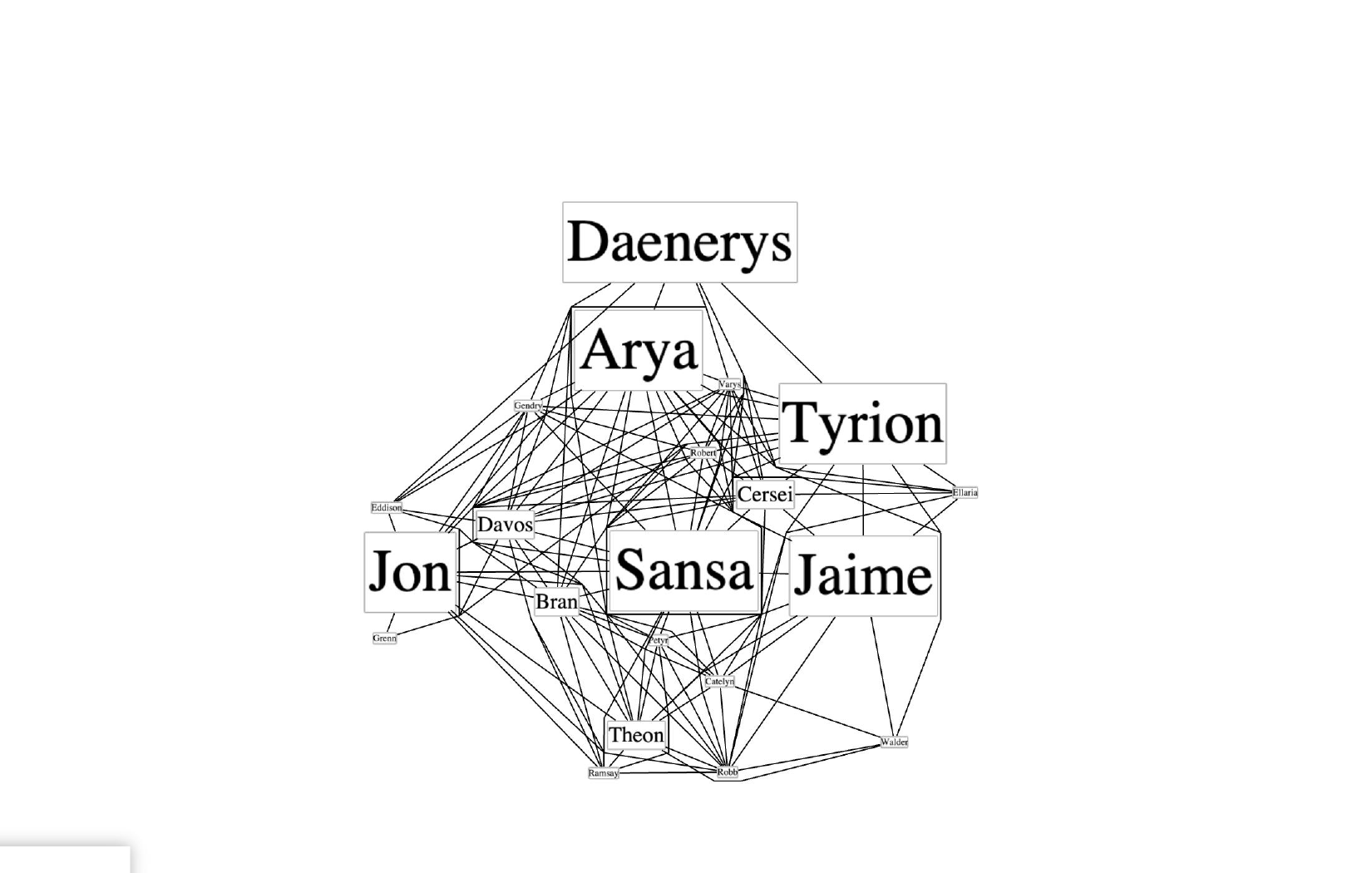}\hfill
  \includegraphics[width=0.48\textwidth]{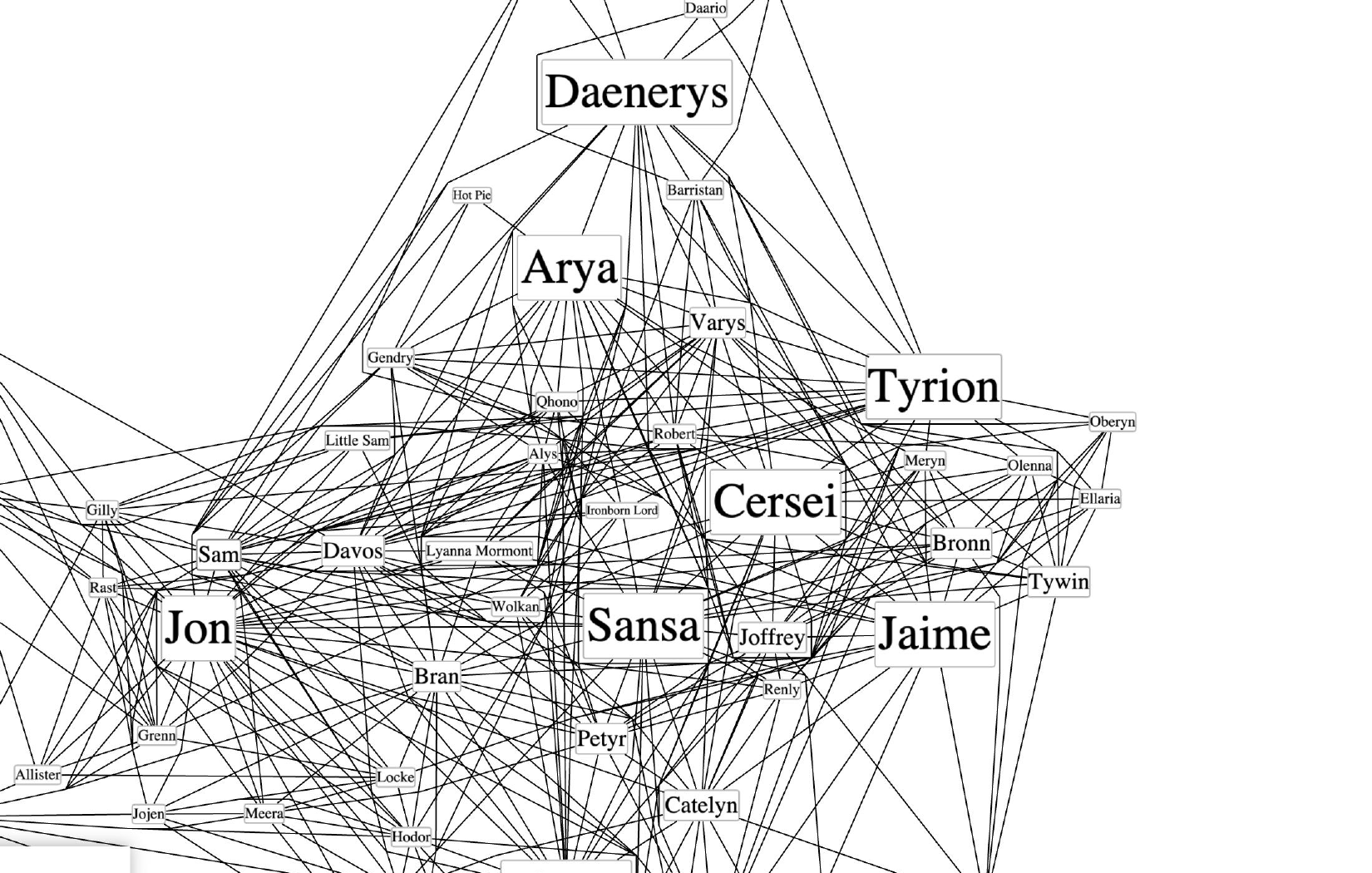}
  \caption{Two adjacent pyramid levels of the Game of Thrones graph rendered by the WebGL renderer of \msagljs{} with the sleeve router. The coarser level is on the left, the finer one on the right; both panels are cropped to the same world window. For each level a new graph is assembled from a PageRank-ranked prefix of the nodes, and edges are routed against the obstacles introduced at that level.}
  \label{fig:tile-pyramid}
\end{figure}

We start from $Z=0$ and grow the pyramid one level at a time. We
stop as soon as any of three conditions is met:
\begin{enumerate}
\item every tile at the new level contains at most $\mathcal{C}$
elements, with default $\mathcal{C}=500$;
\item the new tile width and height are both below ten times the
average node width and height;
\item while generating the tiles of the new finest level, the running
total of stored tile elements summed over all levels, multiplied by
$200$~bytes per element, exceeds the memory budget $\mathcal{M}_{\max}$,
with default $4$~GB; the partial level is then discarded and $Z$ is
set to the previous finest level.
\end{enumerate}
The third condition is checked incrementally inside the level-build
loop, so we abort as soon as the memory barrier is reached. On every
benchmark graph in Table~\ref{tab:max-per-tile} the first condition
fires; the second and third are defensive guards that bound depth on
hypothetical inputs whose finest level cannot be packed within
$\mathcal{C}=500$.
\paragraph{Three phases.} Once $Z$ and the tile rectangles are fixed,
each coarser level $z=Z{-}1, Z{-}2, \dots, 0$ is built in three phases:
\begin{enumerate}
\item pick the nodes of level~$z$ with PageRank and possibly enlarge
  them, Section~\ref{sec:filter};
\item reroute the edges of $G_z$ around the resulting obstacles,
  Section~\ref{sec:sleeve-routing};
\item recompute the per-tile edge clips on level $z$,
  Section~\ref{sec:tile-build}.
\end{enumerate}
Figure~\ref{fig:tile-pyramid} shows two adjacent pyramid levels built
by this procedure; Algorithm~\ref{alg:build-pyramid} summarizes the
whole pipeline.

\subsection{PageRank-Guided Level Construction}
\label{sec:filter}
Before filling the level's tiles for rendering we run
PageRank~\cite{page1999pagerank} on~$G$ once, and sort all nodes of
the graph, $V$, by descending rank. Write $k = Z - z$ for the
level's depth below the finest level. The candidates for a level of
depth $k$ are the nodes of the \emph{prefix} of the sorted array of
length $\lceil |V|/2^k\rceil$. The level's edge set, used by
Algorithm~\ref{alg:build-pyramid}, is the subset of $E(G)$ whose
endpoints both lie in this prefix.

Node positions are inherited from the original layout and never move:
across levels we change only each node's display size. 

We add nodes to a level of depth $k$ by walking the level's prefix in
rank-descending order. The top-ranked node is scaled by $2^k$. We keep
the node scales non-increasing, but as large as possible, while
making sure that each next node does not overlap the previously
selected nodes. If even at its original size the candidate's bounding
box would overlap an already-accepted node's scaled box, the candidate
is dropped.

Dropping a relatively high-ranked candidate at a coarse level is a
deliberate compromise: we trade the inclusion of that candidate at
this level for the invariant that the rendered nodes at every level
remain pairwise disjoint. The cost is small in practice. Consider
Arya's node in the \emph{Game of Thrones} graph: as
Figure~\ref{fig:tile-pyramid-3d} shows, it is dropped from the top
layer because a higher-ranked node, Tyrion, has covered the available
region. The user still sees that part of the graph as occupied, and
Arya's individual node appears at a finer level where its box fits.

\paragraph{Spatial-hash overlap test.}
A naive overlap test against every previously accepted node would
dominate the cost of the filter. We instead maintain the accepted
boxes in a uniform spatial hash whose cell size is chosen so that any
accepted box, and any candidate's box at the level's maximum scale,
both fit inside one cell on each axis. Because of this choice, two
such boxes can overlap only if their center cells differ by at most
one step on each axis: testing a candidate therefore reduces to
scanning the accepted boxes registered in the candidate's own cell and
in the eight cells immediately around it.

The hash brings little benefit at the coarsest levels, where the
prefix is short and even a naive scan over the few accepted boxes is
cheap; it pays off at finer levels, where the prefix grows
geometrically and a per-candidate constant-time test replaces a linear
scan over hundreds of already-accepted boxes.  On the layouts we
tested the expected work per candidate is constant, so the overall
cost is dominated by the initial sort by PageRank.  In our experiments
this phase was never a bottleneck: the per-level routing dominates
the build time at every level we measured.

\subsection{Splitting Edges between Tiles}
\label{sec:tile-build}
We describe edge routing later in Section~\ref{sec:sleeve-routing}, but
for the purpose of this paragraph it is enough to know that each curve
is a polyline.

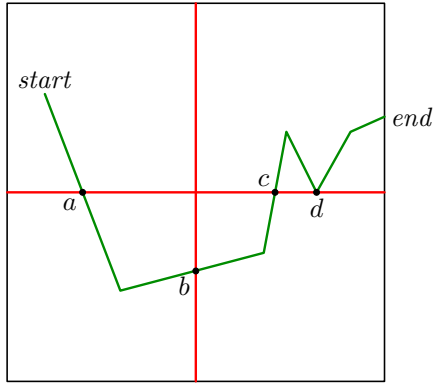
\begin{figure}[!tb]
  \centering
  \begin{tikzpicture}[
      line cap=round,
      line join=round,
      every node/.style={font=\small, inner sep=2pt},
    ]
    \draw[black, line width=0.6pt] (0,0) rectangle (5,5);
    \draw[red, line width=0.9pt] (0,2.5) -- (5,2.5);
    \draw[red, line width=0.9pt] (2.5,0) -- (2.5,5);
    \draw[green!55!black, line width=0.9pt]
         (0.50,3.80)
      -- (1.50,1.20)
      -- (3.40,1.70)
      -- (3.70,3.30)
      -- (4.10,2.50)
      -- (4.55,3.30)
      -- (5.00,3.50);
    \foreach \p in {%
        (1.00,2.50),(2.50,1.46),(3.55,2.50),(4.10,2.50)}
      \filldraw[black] \p circle (1.1pt);
    \node[anchor=south]      at (0.50,3.80) {\emph{start}};
    \node[anchor=north east] at (1.00,2.50) {$a$};
    \node[anchor=north east] at (2.50,1.46) {$b$};
    \node[anchor=south east] at (3.55,2.50) {$c$};
    \node[anchor=north]      at (4.10,2.50) {$d$};
    \node[anchor=west]       at (5.00,3.50) {\emph{end}};
  \end{tikzpicture}
  \caption{An edge clip, drawn as a polyline, from \emph{start} to
    \emph{end} inside a tile. The two red midlines split the tile
    into four sub-tiles and meet the polyline at the four
    points $a$, $b$, $c$, $d$; at $d$ the polyline touches the
    horizontal midline without crossing it. Cutting the polyline at
    these points yields five sub-clips $[\mathit{start},a]$,
    $[a,b]$, $[b,c]$, $[c,d]$, $[d,\mathit{end}]$, each contained in
    one sub-tile and meeting that sub-tile's boundary only at its
    endpoints.}
  \label{fig:clip}
\end{figure}

The edge-clip generation works as follows. Consider a tile and an edge
clip whose curve meets the tile boundary only at the curve's
endpoints. Let the \emph{midlines} of a tile be the horizontal and
vertical segments through its center; they split the tile into four
sub-tiles. Cutting the curve with the midlines yields a set of
sub-clips, each contained in one sub-tile, and each preserves the
invariant: its curve meets the sub-tile boundary only at its
endpoints. Because of this invariant, every split reuses the parent
clip's boundary crossings, and only the curve-vs-midline
intersections need to be computed; no full curve-vs-rectangle clipping
is required. Figure~\ref{fig:clip} illustrates one such split.

When we grow the pyramid in line~\ref{algln:split} of
Algorithm~\ref{alg:build-pyramid}, each edge clip is already
contained in its parent tile, so subdivision starts from that
parent's rectangle. When we prepare the tiles for rendering in
line~\ref{algln:rebuild} of Algorithm~\ref{alg:build-pyramid}, we
have the rerouted graph $G_z$, so for each edge of $G_z$ we start
the subdivision from the top level rectangle.

\paragraph{Bundling edge clips.}
To reduce the number of elements rendered in a tile, all clips within
a tile that share the same, or almost the same, pair of endpoints are
bundled into a single clip whose attached edge list accumulates every
contributing edge; the bundled geometry is taken from the first such
clip encountered during splitting, and the choice is consistent within
each level. Table~\ref{tab:max-per-tile} reports the resulting maximum
number of elements rendered in a single tile across the entire pyramid
for a range of graphs.

\begin{table}[t]
\caption{Maximum number of elements rendered in a single tile across
  the whole pyramid for several graphs (capacity $\mathcal{C}=500$;
  $Z$ chosen by the natural stops of Section~\ref{sec:tiling}). The
  capacity~$\mathcal{C}$ only guides the choice of~$Z$; we have no
  tight analytical bound on the actual rendered per-tile element
  count, which is therefore reported empirically. ``Max tile's level''
  is the level on which the maximum tile was found. In the worst
  cases, \texttt{ca-HepPh} and \texttt{ca-CondMat}, the densest tile
  holds 6--8$\times$ the nominal capacity~$\mathcal{C}$.}
\label{tab:max-per-tile}
\centering
\begin{tabular}{lrrrrr}
\toprule
Graph & $|V|$ & $|E|$ & Levels built & Max per tile & Max tile's level \\
\midrule
gameofthrones      &     407 &   2\,639 & 5 &     324 & 4 \\
composers          &  3\,405 &  13\,832 & 8 &     749 & 3 \\
ca-GrQc            &  5\,242 &  28\,968 & 8 &     359 & 4 \\
ca-HepTh           &  9\,877 &  51\,946 & 8 &  1\,836 & 3 \\
facebook\_combined &  4\,039 &  88\,234 & 9 &     928 & 5 \\
ca-HepPh           & 12\,008 & 236\,978 & 9 &  3\,643 & 3 \\
ca-CondMat         & 23\,133 & 186\,878 & 8 &  2\,949 & 3 \\
deezer\_europe     & 28\,281 &  92\,752 & 8 &  2\,402 & 3 \\
delaunay\_n15      & 32\,768 &  98\,274 & 8 &     283 & 7 \\
\bottomrule
\end{tabular}
\end{table}

\section{Sleeve Routing on the CDT}
\label{sec:sleeve-routing}

To prepare for routing given the graph layout, each node is covered by
a padded obstacle polygon. We compute a Constrained Delaunay
Triangulation \cdt{} on the set of these
polygons~\cite{delaunay1934sphere,domiter2008sweep}. The \emph{dual
graph} $D$ of \cdt{} has one node per triangle and one edge for each
pair of triangles sharing a side. The weight of an edge is the
Euclidean distance between the centroids of its triangles.

\subsection{Routing inside a Sleeve}
\label{sec:sleeve}

We split routing an edge into two subproblems: a \emph{combinatorial}
one---which sequence of triangles to cross---and a \emph{geometric}
one---which exact polyline through that sequence is shortest. The
geometric step is the classical funnel
algorithm~\cite{chazelle1982theorem,hershberger1994computing}, which
pulls the path taut inside the polygon formed by the sleeve triangles.

To obtain the sleeve we run a shortest-path search on~$D$, where
triangles inside obstacles other than the source and target obstacles
are excluded. The natural per-edge instantiation is
A*~\cite{hart1968formal} with the straight-line distance to~$t$ as
heuristic. When a node has many incident edges, however, running
independent searches is wasteful. We instead group edges by source~$s$
and run a single Dijkstra computation per source, expanding until all
targets are reached and recovering each sleeve by following parent
pointers. This gives a speedup over A*; for example on Game of Thrones,
with 407 nodes and 2\,639 edges, it cuts routing time by $\sim$23\,\%.
Table~\ref{tab:routing-modes} reports the gain across the full
benchmark suite.

We can improve even more. Since the query edges are undirected
on~$D$, each can be served from either endpoint, so the number of
distinct roots is in turn reducible by reorienting query edges to
minimize that count. Let the \emph{demand graph}~$H$ be the graph
whose vertices are the graph nodes and whose edges are the query
edges still to be routed. A set of roots~$R$ is feasible iff every
demand edge has at least one endpoint in~$R$, that is, $R$ is a
vertex cover of~$H$, and the smallest such~$R$ is a minimum vertex
cover of~$H$. The problem of finding a minimum vertex cover is
NP-hard~\cite{karp1972reducibility}, so we approximate it
with the standard greedy maximum-degree rule: while $H$ still has an
edge, pick a vertex of maximum degree in~$H$, add it to~$R$, and
delete it together with its incident edges. Implemented with degree
buckets, the procedure runs in linear time. Each demand edge is then
routed from whichever endpoint lies in~$R$. To our knowledge no
prior edge router exploits a vertex cover of the demand graph in
this way; many-to-many shortest-path
frameworks~\cite{knopp2007many} also amortize per-source searches,
but they take the source set as given and do not reduce it through a
covering argument. Applying this heuristic reduces the number of
Dijkstra trees on Game of Thrones from~331, one per distinct source,
to~198, and on the composers graph with 3\,405 nodes and 13\,832
edges from~2\,645 to~1\,281, shortening the routing pass by
$\sim$28\,\% and $\sim$31\,\% respectively.

In addition, before launching the sleeve search for an edge we try routing the
straight line segment between the source and the target by walking the
CDT.

\paragraph{Route quality.}
To assess route quality, we compared the sleeve router with the
optimum on the full visibility graph of the padded obstacles, on the
Game of Thrones layout. The visibility-graph optimum is the shortest
taut polyline around the obstacles; we computed it by all-pairs
Dijkstra over the polygon corners and the node centers. This is a
lower bound on any obstacle-avoiding route. On the $407$-node,
$2{,}639$-edge layout the sleeve router has total arc length
$714.3$\,k arc-length units against $695.1$\,k for the
visibility-graph optimum, a ratio of $1.028\times$. The worst
per-edge ratio is $1.37$, much less than the constrained Delaunay
triangulation's worst-case stretch $4\pi\sqrt{3}/9 \approx
2.42$~\cite{bose2006stretch}.

\paragraph{Edge hint.}
We explored a routing variant in which the path generated for an edge
on one level provides a hint for routing the same edge on an adjacent
level, increasing the visual stability of the routes between
levels. Its shortcoming is that the search cannot be batched per
source as the Dijkstra-tree variant above is, which impacts
performance.

\subsection{Vertex Collapse at Source and Target}
\label{sec:collapse}

\begin{figure}[t]
  \centering
  \begin{minipage}[b]{0.42\textwidth}
    \centering
    \includegraphics[width=\linewidth]{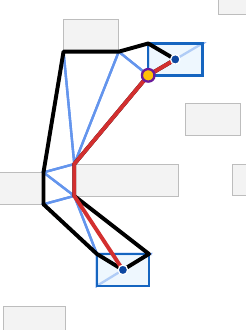}\\[2pt]
    {\small (a) before collapse}
  \end{minipage}\hfill
  \begin{minipage}[b]{0.42\textwidth}
    \centering
    \includegraphics[width=\linewidth]{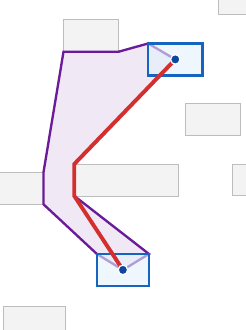}\\[2pt]
    {\small (b) after collapse}
  \end{minipage}
  \caption{Effect of vertex collapse on the LORAS--RENLY edge of the Game of Thrones graph. Endpoints are shown as blue dots; their padded obstacles are highlighted in light blue. (a)~The CDT triangles forming the sleeve are drawn in cornflower blue and the funnel path in red; the thick black polyline is the perimeter of the non-collapsed sleeve. The yellow dot marks the source-obstacle vertex that gets collapsed. (b)~The sleeve after collapse, drawn as a single purple polygon with the source/target obstacles merged into the corresponding endpoints, and the funnel path (red) through it. Collapsing widens the funnel opening at the endpoint and removes the spurious detour around the obstacle corner.}
  \label{fig:collapse}
\end{figure}

We improve the sleeve's geometry in the situation depicted in
Figure~\ref{fig:collapse}. The remedy is to \emph{collapse}
corner-hugging chain vertices to the corresponding endpoint. Walking
the right chain of the sleeve outward from~$s$, we
find the first vertex of the source obstacle at which the chain makes
a right turn, and replace every source-owned vertex on that chain
up to and including this one by~$s$. The left chain is processed
mirror-symmetrically, with left turns instead of right, and the
target end likewise. Diagonals whose two endpoints coincide after
collapse are dropped, and the left/right orientation of each surviving
diagonal is preserved from the uncollapsed geometry. The collapsed
vertex in Figure~\ref{fig:collapse}(a) is marked in yellow.

After collapse, the funnel sees a wide opening at each endpoint, as in
Figure~\ref{fig:collapse}(b), and naturally selects the optimal exit
direction. A final arrowhead-trimming step clips the path at the node
boundary curves. Collapse might cause the resulting path to overlap
nodes other than the source and the target, but we have found the
effect to be negligible.

The implementation also offers an optional pass that fits a Bezier segment into each polyline corner to render edges as smooth curves. It is disabled by default for performance: the polyline rendering is fast and visually acceptable, so we trade the smoother curves for routing speed.

\section{Rendering with deck.gl}
\label{sec:rendering}

The output of the build pipeline of Section~\ref{sec:tiling} is
consumed in the browser by the \texttt{TileLayer} of
\texttt{@deck.gl/geo-layers}~\cite{deckglTileLayer}. \texttt{TileLayer}
is the standard component used by online tile-map renderers: it
tracks the viewport, determines which tiles are needed at the current
zoom, and asynchronously requests them through a user-supplied
\texttt{getTileData} callback, caching the results across frames so
that subsequent pans within the cache require no further work.

\msagljs{} integrates with this component at construction time by
exposing the precomputed pyramid as the \texttt{TileLayer}'s zoom
range, with the coarsest pyramid level mapped to the layer's lowest
zoom and the finest level to the highest. As the user pans and zooms,
\texttt{TileLayer} resolves the viewport to an integer zoom and a set
of axis-aligned tile indices and, for each tile it needs to display,
calls \texttt{getTileData}. The renderer translates each such request
back into a tile of the corresponding pyramid level and returns its
precomputed nodes, edge clips, edge labels, and arrowheads. The
mapping is one-to-one, so no geometric work is performed at view
time: panning shifts the visible window over the tile cache, and
zooming displays tiles from the adjacent level using
\texttt{TileLayer}'s built-in cross-fade.

The rendered geometry depends only on the zoom level and does not
change with the viewport, except for the clipping to the visible
window. Within each level, the displayed nodes are the ones chosen
for that level by the construction of Section~\ref{sec:filter}, and
the displayed edges are the geodesics around those nodes,
computed by the sleeve router of Section~\ref{sec:sleeve-routing}.

\paragraph{Interaction.}
The renderer supports pan, zoom, and hover highlighting of nodes and
edges. Hovering an edge highlights it and its incident nodes; hovering
a node highlights the node, its one-hop neighbors, and its two-hop
neighbors, each in a different color. A \texttt{zoomTo} API animates
the viewport to a target rectangle with a smooth transition and is
used, for example, by the search box to focus the view on a query
result.
\section{Experimental Results}
\label{sec:experiments}

We evaluate the sleeve router and the tile-pyramid build on the same benchmark graphs as Table~\ref{tab:max-per-tile}: the Game of Thrones social network~\cite{beveridge2018game}, the GD~2011 contest \texttt{composers} graph~\cite{composers}, four SNAP collaboration networks (\texttt{ca-GrQc}, \texttt{ca-HepTh}, \texttt{ca-HepPh}, \texttt{ca-CondMat}) and the SNAP \texttt{facebook\_combined} ego network~\cite{leskovec2007graph,fb}, the \texttt{deezer\_europe} social network~\cite{feather}, and the SuiteSparse Delaunay mesh \texttt{delaunay\_n15}. All experiments run in headless Chrome~148 on an Apple~M4~Max laptop with 48\,GB of unified memory, with the unmodified browser bundle of \msagljs{} loaded into the page. Layout uses IPSep-CoLa, and the tile pyramid is built with the same parameters as Table~\ref{tab:max-per-tile}: capacity~$\mathcal{C}=500$, with $Z$ chosen by the natural stops of Section~\ref{sec:tiling}.

Table~\ref{tab:perf} reports per-graph timings of three pipeline stages: \emph{Routing}, the sum of the two phases logged by the sleeve router on the full graph---constrained Delaunay triangulation of the obstacles and the Dijkstra-tree search on the CDT dual, Section~\ref{sec:sleeve}; \emph{Tiling}, the build of the tile pyramid, Section~\ref{sec:tiling}, which includes the per-level routing; and \emph{Total}, the end-to-end loading time including parsing.

\begin{table}[t]
  \centering
  \scriptsize
  \setlength{\tabcolsep}{6pt}
  \begin{tabular}{@{}lrrrrr@{}}
    \toprule
    Graph & $|V|$ & $|E|$ & Routing & Tiling & Total \\
    \midrule
    gameofthrones      &     407 &   2\,639 &   0.07 &   0.07 &   0.38 \\
    composers          &  3\,405 &  13\,832 &   1.49 &   0.74 &   2.69 \\
    ca-GrQc            &  5\,242 &  28\,968 &   1.78 &   0.87 &   3.40 \\
    facebook\_combined &  4\,039 &  88\,234 &   4.34 &   2.36 &   7.77 \\
    ca-HepTh           &  9\,877 &  51\,946 &   9.33 &   3.41 &  14.87 \\
    ca-HepPh           & 12\,008 & 236\,978 &  38.89 &  14.08 &  57.71 \\
    ca-CondMat         & 23\,133 & 186\,878 &  79.72 &  20.52 & 110.06 \\
    deezer\_europe     & 28\,281 &  92\,752 & 134.78 &  23.09 & 172.68 \\
    delaunay\_n15      & 32\,768 &  98\,274 &  18.71 &   5.64 &  41.38 \\
    \bottomrule
  \end{tabular}
  \caption{End-to-end loading benchmarks in headless Chrome~148 on an
    Apple~M4~Max laptop, all times in seconds. Measurements come from
    the \texttt{chrome-routing-bench} example, which runs the
    unmodified browser bundle of \msagljs{}. \emph{Routing} sums the
    CDT-construction and Dijkstra-tree sleeve-search phases of
    Section~\ref{sec:sleeve} on the full graph; \emph{Tiling} is the
    build of the tile pyramid, capacity $\mathcal{C}=500$ with $Z$
    chosen by the natural stops of Section~\ref{sec:tiling}, which
    includes the per-level routing; \emph{Total} is the total
    loading time including parsing of the graph file, graph layout,
    and all other stages.}
  \label{tab:perf}
\end{table}

The CDT phase contributes negligibly to the routing column---from 6\,ms on Game of Thrones to about half a second on the largest graphs---so virtually all of \emph{Routing} is the Dijkstra-tree sleeve search, with the vertex-cover heuristic of Section~\ref{sec:sleeve} reducing the number of trees we run. Even on graphs with hundreds of thousands of edges (\texttt{ca-CondMat}, \texttt{ca-HepPh}, \texttt{deezer\_europe}) the full pipeline finishes in a few minutes; this is a one-time build cost. The resulting tile pyramid already contains the per-level routes, so panning and zooming only fetch and render the precomputed tiles, with no rerouting at view time.

Table~\ref{tab:routing-modes} isolates the contribution of the two
source-side optimizations of Section~\ref{sec:sleeve}: batched
Dijkstra trees over per-edge A* and the vertex-cover heuristic on top
of that.  Each row times a single sleeve-search pass run after layout,
with the constrained Delaunay triangulation excluded so that only the
search costs are compared. The batched Dijkstra trees outperform
per-edge A* on every social graph, with the speedup growing with the
average degree and reaching $+63$\,\% on \texttt{facebook\_combined}
and $+48$\,\% on \texttt{ca-HepPh}; the vertex-cover heuristic yields
a further $11$--$42$\,\% reduction on the same graphs. The Delaunay
mesh \texttt{delaunay\_n15} is the only outlier: the degrees of the
nodes are small and the shortest paths do not extensively overlap,
so neither optimization has a chance to improve performance.

\begin{table}[t]
  \centering
  \scriptsize
  \setlength{\tabcolsep}{4.5pt}
  \begin{tabular}{@{}lrrrrrrr@{}}
    \toprule
    & & & \multicolumn{3}{c}{Time (s)} & \multicolumn{2}{c}{Speedup} \\
    \cmidrule(lr){4-6}\cmidrule(l){7-8}
    Graph & Srcs & VC & A* & DJ & VC\_DJ & DJ/A* & VC/DJ \\
    \midrule
    gameofthrones      &     331 &     198 &   0.077 &   0.059 &   0.043 & $+23$\,\% & $+28$\,\% \\
    composers          &  2\,645 &  1\,281 &   2.60  &   2.14  &   1.49  & $+18$\,\% & $+31$\,\% \\
    ca-GrQc            &  5\,241 &  2\,796 &   2.55  &   2.52  &   1.65  &  $+1$\,\% & $+35$\,\% \\
    facebook\_combined &  3\,663 &  3\,045 &  13.31  &   4.88  &   4.36  & $+63$\,\% & $+11$\,\% \\
    ca-HepTh           &  9\,875 &  5\,003 &  17.13  &  15.42  &   9.28  & $+10$\,\% & $+40$\,\% \\
    ca-HepPh           & 12\,006 &  7\,032 & 111.2   &  58.07  &  36.38  & $+48$\,\% & $+37$\,\% \\
    ca-CondMat         & 23\,133 & 13\,564 & 189.1   & 131.1   &  76.45  & $+31$\,\% & $+42$\,\% \\
    deezer\_europe     & 21\,060 & 13\,450 & 172.1   & 154.0   & 124.4   & $+11$\,\% & $+19$\,\% \\
    delaunay\_n15      & 28\,467 & 23\,456 &  13.57  &  16.02  &  14.88  & $-18$\,\% &  $+7$\,\% \\
    \bottomrule
  \end{tabular}
  \caption{Sleeve-search times of the three routing modes on the
    benchmark graphs of Table~\ref{tab:perf}, in headless
    Chrome~148. Columns: \emph{Srcs}, the number of distinct edge
    endpoints in the demand graph---one per source for the per-edge
    A* baseline; \emph{VC}, the number of Dijkstra trees actually
    launched after the greedy maximum-degree vertex-cover
    reorientation of Section~\ref{sec:sleeve}; \emph{A*}, wall-clock
    seconds of one full sleeve-search pass with per-edge A*\
    (straight-line distance to the target as heuristic);
    \emph{DJ}, wall-clock seconds with the batched Dijkstra-tree
    mode---one tree per source, expanding until all of that source's
    targets are reached; \emph{VC\_DJ}, wall-clock seconds with the
    vertex-cover-reoriented Dijkstra-tree mode---same as DJ but with
    only \emph{VC}~roots; \emph{DJ/A*}~$=$~$(\text{A*}-\text{DJ})/\text{A*}$,
    speedup of DJ over A*; \emph{VC/DJ}~$=$~$(\text{DJ}-\text{VC\_DJ})/\text{DJ}$,
    additional speedup of VC\_DJ over DJ. The CDT-construction phase,
    identical across modes, is excluded so that only the search costs
    are compared.}
  \label{tab:routing-modes}
\end{table}

\paragraph{Browsing smoothness.}
We test how smoothly a user can browse a freshly loaded graph in the
browser. On the same M4~Max laptop as Tables~\ref{tab:perf} and
\ref{tab:routing-modes}, a script loads a graph, waits for the tile
pyramid to build, and then plays back twelve dives. Each dive picks
a random point inside the laid-out graph, recenters the view on that
point, zooms from the coarsest view all the way in to the finest
detail in one second, then zooms back out to the coarsest view in
another second. The twelve random points are independent and the
dives run back to back.

While the dives play, the script measures three things on the
browser side. First, how often the screen actually refreshes: at
$60$\,Hz the screen refreshes every $16.7$\,ms, and a longer gap is
visible as a stutter. We report the $95$th percentile of the gap
between successive refreshes, the value below which $95\,\%$ of
refreshes fall. Second, how often the browser pauses for more than
$50$\,ms on a single block of work; this is the standard browser
metric of a ``long task'' and each one freezes the page. Third, at
the deepest point of every dive, how many graphical objects the
browser must draw, that is, nodes, edge clips, labels, and
arrowheads, summed over the tiles that fall inside the viewport.
This last number is a proxy for how hard the GPU has to work each
frame. Table~\ref{tab:smoothness} reports the result.

\begin{table}[t]
  \centering
  \scriptsize
  \setlength{\tabcolsep}{4pt}
  \begin{tabular}{@{}lrrrrrrr@{}}
    \toprule
    Graph & $|V|$ & $|E|$ & $Z$ & LT & LT total (ms) & Frame p95 (ms) & Elts/frame max \\
    \midrule
    gameofthrones      &     407 &   2\,639 & 5 &  0 &     0 &  17.4 &   2\,567 \\
    composers          &  3\,405 &  13\,832 & 9 &  1 &    51 &  18.2 &      497 \\
    ca-GrQc            &  5\,242 &  28\,968 & 9 &  0 &     0 &  18.2 &      677 \\
    ca-HepTh           &  9\,877 &  51\,946 & 9 &  0 &     0 &  18.2 &      383 \\
    facebook\_combined &  4\,039 &  88\,234 & 9 &  0 &     0 &  18.0 &      495 \\
    ca-HepPh           & 12\,008 & 236\,978 & 9 &  9 &   556 &  18.4 &   1\,394 \\
    ca-CondMat         & 23\,133 & 186\,878 & 9 &  5 &   273 &  18.2 &   1\,391 \\
    deezer\_europe     & 28\,281 &  92\,752 & 9 &  3 &   164 &  18.4 &   2\,505 \\
    delaunay\_n15      & 32\,768 &  98\,274 & 9 &  0 &     0 &  18.1 &   1\,012 \\
    \bottomrule
  \end{tabular}
  \caption{Browsing-smoothness measurements on the benchmark graphs.
    For each graph the script plays back twelve dives; each dive picks
    a random point inside the laid-out graph, recenters on it, and
    zooms from the coarsest view to the finest and back in two
    seconds. \emph{$Z$}, number of tile-pyramid levels built;
    \emph{LT}, number of main-thread pauses longer than $50$\,ms
    observed during the twelve dives; \emph{LT total}, the total
    duration of those pauses, in milliseconds; \emph{Frame p95}, the
    $95$th percentile of the gap between successive on-screen
    refreshes during the twelve dives, in milliseconds;
    \emph{Elts/frame max}, the largest number of graphical objects,
    that is, nodes, edge clips, labels, and arrowheads, that the
    browser had to draw at the deepest point of any dive.}
  \label{tab:smoothness}
\end{table}

Across all nine graphs the dives stay close to the $60$\,Hz target.
On every graph, $95\,\%$ of the gaps between successive refreshes
are at most $18.4$\,ms, only $1.7$\,ms over the ideal $16.7$\,ms gap
at $60$\,Hz and too small to be perceived as a stutter. The
browser pauses for more than $50$\,ms only a handful of times across
the twelve dives on the three densest graphs, \texttt{ca-HepPh},
\texttt{ca-CondMat}, and \texttt{deezer\_europe}, and not at all on
the smaller ones, so the moments of stutter that do happen are short
and rare. At the deepest zoom, the browser is asked to draw at most
$2\,567$ objects on \texttt{gameofthrones} and at most $2\,505$ on
\texttt{deezer\_europe} even though the latter graph has $35$ times
more edges. This is the property the pyramid is designed to deliver:
how hard the GPU has to work each frame depends on what is currently
on screen, not on the total size of the graph.

\section{Conclusion}
\label{sec:conclusion}

We presented \msagljs{}, an open-source TypeScript library for graph layout, edge routing, and visualization in web browsers. The sleeve routing method routes edges directly on the CDT dual graph using batched Dijkstra trees and the funnel algorithm. The tiling scheme enables map-like exploration of large graphs with level-of-detail filtering and edge simplification. The WebGL renderer, built on deck.gl, provides smooth pan-and-zoom interaction.

\paragraph{Future work.} The current approach builds the entire tile
pyramid up front, which is expensive and bounds the size of graphs we
can handle. A natural next step is to build only the first few levels
quickly and hand them to the renderer; the remaining levels would be
built lazily as the user browses the graph. This
would lower the time-to-first-frame and lift the limit on graph size.

A second challenge is that some tiles can contain many more
elements than the capacity parameter~$\mathcal{C}$ that guides the
build. Table~\ref{tab:max-per-tile} reports densest tiles holding
6--8$\times$ the nominal $\mathcal{C}$ in the worst cases, which leaves
the rendering and interaction cost uneven across the viewport.
Enforcing an actual per-tile budget remains an open problem.

\bibliography{main}

\begin{thebibliography}{10}

\bibitem{cosmograph}
Cosmograph.
\newblock \url{https://cosmograph.app}.

\bibitem{deckgl}
deck.gl: Large-scale {WebGL}-powered data visualization.
\newblock \url{https://deck.gl/}.

\bibitem{fb}
facebookcombined.
\newblock \url{https://snap.stanford.edu/data/facebook\_combined.txt.gz}.

\bibitem{pathOpt}
Funnel algorithm.
\newblock \url{https://page.mi.fu-berlin.de/mulzer/notes/alggeo/polySP.pdf}.

\bibitem{graphviz}
Graphviz.
\newblock \url{http://www.graphviz.org/}.

\bibitem{regraph}
Regraph.
\newblock \url{https://cambridge-intelligence.com/regraph/}.

\bibitem{sfdp}
sfdp.
\newblock \url{https://graphviz.org/docs/layouts/sfdp/}.

\bibitem{sigmajs}
Sigma.js: a {JavaScript} library aimed at visualizing graphs of thousands of
  nodes and edges.
\newblock \url{https://www.sigmajs.org/}.

\bibitem{composers}
Skewed.
\newblock
  \url{http://mozart.diei.unipg.it/gdcontest/contest2011/composers.xml}.

\bibitem{deckglTileLayer}
{TileLayer} ({\tt @deck.gl/geo-layers}).
\newblock \url{https://deck.gl/docs/api-reference/geo-layers/tile-layer}.
\newblock Accessed: 2026.

\bibitem{yworks}
yworks.
\newblock \url{https://yworks.com/products/yed}.

\bibitem{abello2006askgraphview}
James Abello, Frank Van~Ham, and Neeraj Krishnan.
\newblock {ASK-GraphView}: A large scale graph visualization system.
\newblock In {\em IEEE Transactions on Visualization and Computer Graphics},
  volume~12, pages 669--676. IEEE, 2006.

\bibitem{beveridge2018game}
Andrew Beveridge and Michael Chemers.
\newblock The game of game of thrones: Networked concordances and fractal
  dramaturgy.
\newblock In {\em Reading Contemporary Serial Television Universes}, pages
  201--225. Routledge, 2018.

\bibitem{bose2006stretch}
Prosenjit Bose and J.~Mark Keil.
\newblock On the stretch factor of the constrained {Delaunay} triangulation.
\newblock In {\em 3rd International Symposium on Voronoi Diagrams in Science
  and Engineering (ISVD)}, pages 25--31. IEEE, 2006.

\bibitem{brandes2007eigensolver}
Ulrik Brandes and Christian Pich.
\newblock Eigensolver methods for progressive multidimensional scaling of large
  data.
\newblock In {\em Graph Drawing: 14th International Symposium, GD 2006,
  Karlsruhe, Germany, September 18-20, 2006. Revised Papers 14}, pages 42--53.
  Springer, 2007.

\bibitem{chazelle1982theorem}
Bernard Chazelle.
\newblock A theorem on polygon cutting with applications.
\newblock In {\em 23rd Annual Symposium on Foundations of Computer Science
  (sfcs 1982)}, pages 339--349. IEEE, 1982.

\bibitem{delaunay1934sphere}
Boris Delaunay et~al.
\newblock Sur la sphere vide.
\newblock {\em Izv. Akad. Nauk SSSR, Otdelenie Matematicheskii i Estestvennyka
  Nauk}, 7(1):793--800, 1934.

\bibitem{demyen2006efficient}
Douglas Demyen and Michael Buro.
\newblock Efficient triangulation-based pathfinding.
\newblock In {\em Proceedings of the 21st National Conference on Artificial
  Intelligence (AAAI)}, pages 942--947, 2006.

\bibitem{dobkin1997implementing}
David~P Dobkin, Emden~R Gansner, Eleftherios Koutsofios, and Stephen~C North.
\newblock Implementing a general-purpose edge router.
\newblock In {\em Graph Drawing: 5th International Symposium, GD'97 Rome,
  Italy, September 18--20, 1997 Proceedings 5}, pages 262--271. Springer, 1997.

\bibitem{domiter2008sweep}
Vid Domiter and Borut {\v{Z}}alik.
\newblock Sweep-line algorithm for constrained delaunay triangulation.
\newblock {\em International Journal of Geographical Information Science},
  22(4):449--462, 2008.

\bibitem{duan2025sorting}
Ran Duan, Jiayi Mao, Xiao Mao, Xinkai Shu, and Longhui Yin.
\newblock Breaking the sorting barrier for directed single-source shortest
  paths.
\newblock In {\em Proceedings of the 57th Annual ACM Symposium on Theory of
  Computing (STOC)}, 2025.
\newblock \url{https://arxiv.org/abs/2504.17033}.

\bibitem{dwyer2006ipsepcola}
Tim Dwyer, Yehuda Koren, and Kim Marriott.
\newblock {IPSep-CoLa}: An incremental procedure for separation constraint
  layout of graphs.
\newblock {\em IEEE Transactions on Visualization and Computer Graphics},
  12(5):821--828, 2006.

\bibitem{dwyer2010fast}
Tim Dwyer and Lev Nachmanson.
\newblock Fast edge-routing for large graphs.
\newblock In {\em Graph Drawing: 17th International Symposium, GD 2009,
  Chicago, IL, USA, September 22-25, 2009. Revised Papers 17}, pages 147--158.
  Springer, 2010.

\bibitem{franz2016cytoscape}
Max Franz, Christian~T. Lopes, Gerardo Huck, Yue Dong, Onur Sumer, and Gary~D.
  Bader.
\newblock {Cytoscape.js}: a graph theory library for visualisation and
  analysis.
\newblock {\em Bioinformatics}, 32(2):309--311, 2016.

\bibitem{gansner2005topological}
Emden~R. Gansner, Yehuda Koren, and Stephen North.
\newblock Topological fisheye views for visualizing large graphs.
\newblock {\em IEEE Transactions on Visualization and Computer Graphics},
  11(4):457--468, 2005.

\bibitem{gansner2000graphviz}
Emden~R. Gansner and Stephen~C. North.
\newblock An open graph visualization system and its applications to software
  engineering.
\newblock {\em Software: Practice and Experience}, 30(11):1203--1233, 2000.

\bibitem{guibas1987linear}
Leonidas Guibas, John Hershberger, Daniel Leven, Micha Sharir, and Robert~E.
  Tarjan.
\newblock Linear-time algorithms for visibility and shortest path problems
  inside triangulated simple polygons.
\newblock {\em Algorithmica}, 2(1--4):209--233, 1987.

\bibitem{harel2002multiscale}
David Harel and Yehuda Koren.
\newblock A fast multi-scale method for drawing large graphs.
\newblock In {\em Journal of Graph Algorithms and Applications}, volume~6,
  pages 179--202, 2002.

\bibitem{hart1968formal}
Peter~E. Hart, Nils~J. Nilsson, and Bertram Raphael.
\newblock A formal basis for the heuristic determination of minimum cost paths.
\newblock {\em IEEE Transactions on Systems Science and Cybernetics},
  4(2):100--107, 1968.

\bibitem{hershberger1994computing}
John Hershberger and Jack Snoeyink.
\newblock Computing minimum length paths of a given homotopy class.
\newblock {\em Computational geometry}, 4(2):63--97, 1994.

\bibitem{hershberger1999optimal}
John Hershberger and Subhash Suri.
\newblock An optimal algorithm for {Euclidean} shortest paths in the plane.
\newblock {\em SIAM Journal on Computing}, 28(6):2215--2256, 1999.

\bibitem{holten2006hierarchical}
Danny Holten.
\newblock Hierarchical edge bundles: Visualization of adjacency relations in
  hierarchical data.
\newblock {\em IEEE Transactions on Visualization and Computer Graphics},
  12(5):741--748, 2006.

\bibitem{holten2009fdeb}
Danny Holten and Jarke~J. Van~Wijk.
\newblock Force-directed edge bundling for graph visualization.
\newblock In {\em Computer Graphics Forum}, volume~28, pages 983--990. Wiley
  Online Library, 2009.

\bibitem{hu2005efficient}
Yifan Hu.
\newblock Efficient, high-quality force-directed graph drawing.
\newblock {\em Mathematica Journal}, 10(1):37--71, 2005.

\bibitem{hurter2012graph}
Christophe Hurter, Ozan Ersoy, and Alexandru Telea.
\newblock Graph bundling by kernel density estimation.
\newblock In {\em Computer graphics forum}, volume~31, pages 865--874. Wiley
  Online Library, 2012.

\bibitem{karp1972reducibility}
Richard~M. Karp.
\newblock Reducibility among combinatorial problems.
\newblock In Raymond~E. Miller, James~W. Thatcher, and Jean~D. Bohlinger,
  editors, {\em Complexity of Computer Computations}, pages 85--103. Plenum
  Press, New York, 1972.

\bibitem{knopp2007many}
Sebastian Knopp, Peter Sanders, Dominik Schultes, Falk Schieferdecker, and
  Dorothea Wagner.
\newblock Computing many-to-many shortest paths using highway hierarchies.
\newblock In {\em Proceedings of the 9th Workshop on Algorithm Engineering and
  Experiments (ALENEX)}, pages 36--45. SIAM, 2007.

\bibitem{lee1984euclidean}
D.~T. Lee and Franco~P. Preparata.
\newblock {Euclidean} shortest paths in the presence of rectilinear barriers.
\newblock {\em Networks}, 14(3):393--410, 1984.

\bibitem{leskovec2007graph}
Jure Leskovec, Jon Kleinberg, and Christos Faloutsos.
\newblock Graph evolution: Densification and shrinking diameters.
\newblock {\em ACM transactions on Knowledge Discovery from Data (TKDD)},
  1(1):2--es, 2007.

\bibitem{nachmanson2015graphmaps}
Lev Nachmanson, Roman Prutkin, Bongshin Lee, Nathalie~Henry Riche, Alexander~E
  Holroyd, and Xiaoji Chen.
\newblock Graphmaps: Browsing large graphs as interactive maps.
\newblock In {\em Graph Drawing and Network Visualization: 23rd International
  Symposium, GD 2015, Los Angeles, CA, USA, September 24-26, 2015, Revised
  Selected Papers 23}, pages 3--15. Springer, 2015.

\bibitem{nachmanson2004glee}
Lev Nachmanson, George~G. Robertson, and Bongshin Lee.
\newblock Drawing graphs with {GLEE}.
\newblock In {\em Graph Drawing: 15th International Symposium, GD 2007}, pages
  389--394. Springer, 2008.

\bibitem{page1999pagerank}
Lawrence Page, Sergey Brin, Rajeev Motwani, and Terry Winograd.
\newblock The pagerank citation ranking: Bringing order to the web.
\newblock {\em Stanford InfoLab}, 249(373):1--17, 1999.

\bibitem{perrot2018cornac}
Alexandre Perrot and David Auber.
\newblock Cornac: Tackling huge graph visualization with big data
  infrastructure.
\newblock {\em IEEE Transactions on Big Data}, 6(1):80--92, 2018.

\bibitem{feather}
Benedek Rozemberczki and Rik Sarkar.
\newblock {Characteristic Functions on Graphs: Birds of a Feather, from
  Statistical Descriptors to Parametric Models}.
\newblock In {\em Proceedings of the 29th ACM International Conference on
  Information and Knowledge Management (CIKM '20)}, page 1325–1334. ACM,
  2020.

\bibitem{shewchuk2002delaunay}
Jonathan~Richard Shewchuk.
\newblock Delaunay refinement algorithms for triangular mesh generation.
\newblock {\em Computational Geometry}, 22(1--3):21--74, 2002.

\bibitem{sugiyama1981}
Kozo Sugiyama, Shojiro Tagawa, and Mitsuhiko Toda.
\newblock Methods for visual understanding of hierarchical system structures.
\newblock {\em IEEE Transactions on Systems, Man, and Cybernetics},
  11(2):109--125, 1981.

\bibitem{wybrow2010orthogonal}
Michael Wybrow, Kim Marriott, and Peter~J. Stuckey.
\newblock Orthogonal connector routing.
\newblock {\em Lecture Notes in Computer Science}, 5849:219--231, 2010.

\end{thebibliography}

\end{document}